\documentstyle[amstex,aps,epsfig]{revtex}
\tightenlines

\begin{document}
\title{Anomalous single top quark production at the THERA and Linac$\otimes $LHC
based $\gamma p$ colliders}
\author{O. \c{C}ak\i r$^{a}$, S. Sultansoy$^{b,c}$, M.Y\i lmaz$^{b}$}
\address{$^{a}$Ankara University, Faculty of Sciences, \\
Department of Physics, 06100, \\
Tandogan, Ankara, Turkey.\\
$^{b}$Gazi University, Faculty of Arts and Sciences,\\
Department of Physics, 06500, \\
Besevler, Ankara, Turkey.\\
$^{c}$Institute of Physics, Academy of Sciences, \\
H. Cavid Avenue, 370143, Baku, Azerbaijan.}
\maketitle

\begin{abstract}
Single production of t-quarks at the THERA and Linac$\otimes $LHC based $%
\gamma p$ colliders via anomalous $\gamma ut$ and $\gamma ct$ couplings have
been studied. We show that $\gamma p$ colliders will be a powerful tool for
searching for the anomalous couplings.
\end{abstract}

Although the standard model (SM) has been proved to be phenomenologically
successful at the available energies, there have been intensive studies to
test the deviations from the SM at higher energy scales. Because of its
large mass ($m_{t}\cong 175$ GeV), the top quark is believed to be more
sensitive to new physics than other particles. Recently, the production of
single t-quarks at LEP and HERA was studied in \cite{1}. A possible
anomalous $\gamma ut$ and $\gamma ct$ couplings are generated in a dynamical
theory of mass generation. These anomalous vertices can be examined at
future lepton and lepton-hadron colliders. An essential step in this
direction will be provided by THERA \cite{2} and Linac$\otimes $LHC \cite{3}
based $\gamma p$ colliders (see also review \cite{4}). The main parameters
of these $\gamma p$ colliders are given in the Table \ref{table1}.

Although CDF \cite{5} have shown that $t\rightarrow cg$ and $t\rightarrow
c\gamma $ decays are not the most significant decay modes, high energy
photon may provide anomalous single top production with the anomalous
couplings accessible in the realistic ranges. In this note we study the
potential of the $\gamma p$ colliders in search for single $t$ quark
production in the resonance channel via anomalous coupling.

The possible anomalous couplings of top quarks lead to the following
effective lagrangian for the neutral current interactions between the
fermions and the gauge bosons

\begin{align}
L^{eff}& =L^{SM}+L^{A} \\
L^{A}& =\frac{g_{e}}{\Lambda }\overline{t}\sigma _{\mu \nu }(A_{\gamma
}+B_{\gamma }\gamma _{5})qF^{\mu \nu }+\frac{g_{Z}}{\Lambda }\overline{t}%
\sigma _{\mu \nu }(A_{Z}+B_{Z}\gamma _{5})qZ^{\mu \nu }  \nonumber \\
& +\frac{g_{s}}{\Lambda }\overline{t}\sigma _{\mu \nu }(A_{g}+B_{g}\gamma
_{5})\frac{\lambda ^{a}}{2}qG_{a}^{\mu \nu }+h.c.
\end{align}
where $F^{\mu \nu },Z^{\mu \nu },$ and $G^{\mu \nu }$ are the field strength
tensors of the photon, Z boson and gluons, respectively; $\lambda $ is the
QCD structure constant; $g_{e},g_{Z},$ and $g_{s}$ are the electroweak, and
strong coupling constants, respectively. Constants $A_{\gamma ,Z,g}$ and $%
B_{\gamma ,Z,g}$ are the parameters for the projection operators and
anomalous couplings. Finally, $\Lambda $ is the cutoff of the effective
theory.

Feynman diagram for single production of $t$ quarks at $\gamma p$ collisions
is shown in Fig. \ref{fig1}. Anomalous interaction of the $t$ quarks with
the photon is given explicitly 
\begin{equation}
L_{\gamma }^{A}=\frac{g_{e}}{\Lambda }\overline{t}\sigma _{\mu \nu }\left[
(A_{u}^{t}+B_{u}^{t}\gamma _{5})u+(A_{c}^{t}+B_{c}^{t}\gamma _{5})c\right]
F^{\mu \nu }+h.c.
\end{equation}
In general, the vertex factor for the anomalous top quark couplings can be
rewritten including all undetermined constants and the scale parameter $%
\Lambda $

\begin{equation}
\frac{i}{m_{t}}(f_{1}+f_{2}\gamma _{5})\sigma ^{\mu \nu }k_{\nu }
\end{equation}
where

\begin{equation}
f_{1}=g_{e}A_{i}\frac{m_{t}}{\Lambda }\text{,\qquad\ }f_{2}=g_{e}B_{i}\frac{%
m_{t}}{\Lambda }
\end{equation}

Decay width for top quarks in the SM channel $t\rightarrow bW$ is well known

\begin{equation}
\Gamma (t\rightarrow qW)=\frac{g_{W}^{2}}{64\pi }\frac{|V_{tb}|^{2}}{%
m_{W}^{2}}m_{t}^{3}\left( 1-\frac{m_{W}^{2}}{m_{t}^{2}}\right) \left( 1+%
\frac{m_{W}^{2}}{m_{t}^{2}}-2\frac{m_{W}^{4}}{m_{t}^{4}}\right)
\end{equation}
which is dominant in the full decay mode. The total decay width may be
enhanced by the anomalous decays. In this case the total decay widths will
be the sum of all possible decay contributions

\begin{equation}
\Gamma (t\rightarrow qg)=\frac{g_{s}^{2}}{8\pi }\frac{C_{F}}{\Lambda ^{2}}%
\left( |A_{g}|^{2}+|B_{g}|^{2}\right) m_{t}^{3}
\end{equation}

\begin{equation}
\Gamma (t\rightarrow q\gamma )=\frac{g_{e}^{2}}{8\pi }\frac{1}{\Lambda ^{2}}%
\left( |A_{\gamma }|^{2}+|B_{\gamma }|^{2}\right) m_{t}^{3}
\end{equation}

\begin{equation}
\Gamma (t\rightarrow qZ)=\frac{g_{Z}^{2}}{8\pi }\frac{1}{\Lambda ^{2}}%
(|A_{Z}|^{2}+|B_{Z}|^{2})m_{t}^{3}\left( 1-\frac{m_{Z}^{2}}{m_{t}^{2}}%
\right) \left( 1-\frac{m_{Z}^{2}}{2m_{t}^{2}}-\frac{m_{Z}^{4}}{2m_{t}^{4}}%
\right)
\end{equation}
where $C_{F}$ is the color factor $4/3.$ Here, neglecting the terms $%
(m_{W,Z}/m_{t})^{2}<1,$ we estimate the ratio of the partial widths in the
various channels $V=\gamma ,Z,g$:

\begin{equation}
R_{V}=\frac{\Gamma (t\rightarrow qV)}{\Gamma (t\rightarrow bW)}\approx \frac{%
8g_{V}^{2}\left( |A_{V}|^{2}+|B_{V}|^{2}\right) m_{W}^{2}}{g_{W}^{2}\Lambda
^{2}|V_{tb}|^{2}}
\end{equation}

If we assume that all the constants $A_{V}$ and $B_{V}$ are of the same
order, then the branchings are simply proportional to the gauge couplings $%
g_{V}$

\begin{equation}
g_{s}\approx 1.\,\allowbreak 12\text{, }g_{e}\approx 0.\,\allowbreak 31\text{%
, }g_{W}\approx 0.64\text{ and }g_{Z}\approx 0.\,\allowbreak 74\text{ at }%
q^{2}=m_{Z}^{2}
\end{equation}
where the dominant channel will be the gluon mediated anomalous interaction.
Experimental limits for the anomalous decay channels of top quarks are given
in \cite{6}:

\begin{equation}
R_{\gamma }(t\rightarrow q\gamma )<0.032\text{ \ and }R_{Z}(t\rightarrow
qZ)<0.33
\end{equation}
and the CDF data \cite{5} for branching ratio of top quark decaying to
bottom quark places the limit on the anomalous decay 
\begin{equation}
BR(t\rightarrow qg)<0.45
\end{equation}

We present the ratio of the partial widths for the top quark anomalous
photonic decay vs. anomalous coupling parameter $f$ in Fig. \ref{fig2}. In
the same Figure, corresponding experimental bound is also given. One can see
that SM channel ($t\rightarrow bW$) is dominant for realistic values of
anomalous coupling $f<0.06$.

The differential cross section for the resonant production of top quarks via
the subprocess $\gamma q\rightarrow t\rightarrow Wb$ is

\begin{eqnarray}
\frac{d\widehat{\sigma }}{d\widehat{t}} &=&\frac{g_{W}^{2}|V_{tb}|^{2}}{%
64\pi }\left[ \frac{(A-B)^{2}\left[ m_{t}^{2}\left( 2m_{W}^{4}+\widehat{s}%
\widehat{t}-2m_{W}^{2}(\widehat{s}+\widehat{t})\right) \right] }{%
m_{t}^{2}m_{W}^{2}\widehat{s}\left[ (\widehat{s}-m_{t}^{2})^{2}+m_{t}^{2}%
\Gamma _{t}^{2}\right] }\right.  \nonumber \\
&&-\left. \frac{\left[ \widehat{s}^{2}(\widehat{s}+\widehat{t})-m_{W}^{2}%
\widehat{s}(\widehat{s}+2\widehat{t})\right] (A+B)^{2}}{m_{t}^{2}m_{W}^{2}%
\widehat{s}\left[ (\widehat{s}-m_{t}^{2})^{2}+m_{t}^{2}\Gamma _{t}^{2}\right]
}\right]
\end{eqnarray}
corresponding cross section is given by

\begin{eqnarray}
\widehat{\sigma }(\gamma q &\rightarrow &t\rightarrow Wb)=\frac{%
g_{W}^{2}|V_{tb}|^{2}}{128\pi }\frac{\left( \widehat{s}-m_{W}^{2}\right)
^{2}\left( \widehat{s}+2m_{W}^{2}\right) }{m_{t}^{2}m_{W}^{2}}  \nonumber \\
&&\times \frac{\left[ (A^{2}+B{}^{2})(\widehat{s}+m_{t}^{2})-2AB(\widehat{s}%
-m_{t}^{2})\right] }{\widehat{s}\left[ (\widehat{s}-m_{t}^{2})^{2}+m_{t}^{2}%
\Gamma _{t}^{2}\right] }
\end{eqnarray}

In order to see how the anomalous coupling parameters $f_{i}$ change the
transverse momentum distributions of the quark-jet, we derive the following
formula ($V=W,Z,\gamma $)

\begin{eqnarray}
\frac{d\sigma }{dp_{T}}(\gamma p &\rightarrow &V+jet)=2p_{T}\int_{y_{\min
}}^{y^{\max }}dy\int_{x_{a}^{\min }}^{0.83}dx_{a}f_{\gamma
/e}(x_{a})f_{q/p}(x_{b},Q_{p}^{2})  \nonumber \\
&&\times \frac{x_{a}x_{b}s}{x_{a}s-2m_{T}E_{p}e^{y}}\frac{d\widehat{\sigma }%
}{d\widehat{t}}(\widehat{s},\widehat{t},f_{i})
\end{eqnarray}
where

\begin{eqnarray}
y_{\min }^{(\max )} &=&\log \left[ x\pm \sqrt{x^{2}-0.83E_{e}/E_{p}}\right]
\\
x &=&\frac{0.83s+m_{q}^{2}-m_{V}^{2}}{4m_{T}E_{p}},\quad
m_{T}=m_{q}^{2}+p_{T}^{2}
\end{eqnarray}

\begin{eqnarray}
x_{a}^{\min } &=&\max (x_{a}^{(1)},x_{a}^{(2)}) \\
x_{a}^{(1)} &=&\frac{2m_{T}E_{p}e^{y}-m_{q}^{2}+m_{V}^{2}}{%
s-2m_{T}E_{e}e^{-y}},\quad x_{a}^{(2)}=\frac{(m_{q}+m_{V})^{2}}{s} \\
x_{b} &=&\frac{2m_{T}E_{e}x_{a}e^{-y}-m_{q}^{2}+m_{V}^{2}}{%
x_{a}s-2m_{T}E_{p}e^{y}}
\end{eqnarray}
with the Mandelstam variables

\begin{equation}
\widehat{s}=x_{a}x_{b}s,\quad \widehat{t}=m_{q}^{2}-2E_{e}x_{a}m_{T}e^{-y}
\end{equation}
For the numerical calculation, we have used the quark distributions $%
f_{q/p}(x_{b},Q_{p}^{2})$ \cite{7} in the proton and the Compton
backscattered high energy photon spectrum $f_{\gamma /e}(x_{a})$ \cite{8}

\begin{equation}
f_{\gamma /e}(x)=\left\{ 
\begin{array}{c}
N\left[ 1-x+\frac{1}{1-x}\left[ 1-\frac{4x}{x_{0}}(1-\frac{x}{x_{0}(1-x)})%
\right] \right] , \\ 
\multicolumn{1}{l}{0,}
\end{array}
\begin{array}{c}
0<x<x_{\max } \\ 
x>x_{\max }
\end{array}
\right.  \label{Eq.23}
\end{equation}
where $x_{0}=4.82,$ $x_{\max }=x_{0}/(1+x_{0}),$ $N=1/1.84$. The $p_{T}$
distributions of the b-jet in the final state for THERA 1,2 and 3 options,
and Linac$\otimes $LHC are shown in Figures \ref{fig3}-\ref{fig5}.

The differential cross sections for the signal processes $\gamma
u\rightarrow t\rightarrow Wb$ and $\gamma c\rightarrow t\rightarrow Wb$ via
transverse momentum of the $b-$jet are peaked around the value

\begin{equation}
p_{T}^{b}=\sqrt{\left[ \frac{m_{t}^{2}-m_{W}^{2}+m_{b}^{2}}{2m_{t}}\right]
^{2}-m_{b}^{2}}\approx 69\text{ GeV}
\end{equation}
whereas the backgrounds contribute mainly ($\sim 2\times 10^{-4}$ pb/GeV) at
low $p_{T}.$ The cross sections for higher values of the anomalous couplings
show up over the background continuum.

In the case of the Option 1(or 2) of the THERA, about the $70\%$ of the
signal cross section ($\Delta \sigma ^{S}\sim 12.4$ fb at $f=10^{-3})$ lies
in the $p_{T}$ window $50-70$ GeV. Whereas the background cross section in
this interval is only be $\Delta \sigma ^{B}\sim 0.9$ fb. Therefore, high $%
p_{T}$ cut in this interval could help to eliminate background from the
signal. The values of the cross sections for both signal and backgrounds for
different $f$ can be found in Table \ref{table2}. Here, the differential
cross sections are integrated over a chosen $p_{T}$ window (50-70 GeV) in
order to find the statistical significance S$/\sqrt{\text{S+B}}$ (here S
stands for signal and B for background). From the Table \ref{table2}, one
can see that anomalous couplings down to $10^{-3}$ can be reached at THERA
based $\gamma p$ collider. This value could be $10^{-4}$ at Linac$\otimes $%
LHC based $\gamma p$ collider. We have used the integrated luminosity for
the THERA options as $L_{1}^{int}=120$pb$^{-1},$ $L_{2}^{int}=750$pb$^{-1},$ 
$L_{3}^{int}=480$pb$^{-1}$ and the luminosity $L^{int}=3\times 10^{4}$pb$%
^{-1}$ for Linac$\otimes $LHC based $\gamma p$ collider.

Let us estimate the total cross section for the signal process $\gamma
p\rightarrow WbX$, via the resonant production of top quark. The signal and
background total cross sections can be written as

\begin{equation}
\sigma (\gamma p\rightarrow WbX)=\int_{\tau _{\min }}^{0.83}d\tau \int_{\tau
/0.83}^{1}\frac{dx}{x}f_{\gamma /e}(\tau /x)f_{q/p}(x,Q_{p}^{2})\widehat{%
\sigma }(\tau s,f_{i})  \label{Eq.25}
\end{equation}

The total cross sections for the single top production $\gamma p\rightarrow
t\rightarrow Wb$ depending on the coupling $f$ and the background $\gamma
p\rightarrow Wb$ using laser and WW photon spectrum at THERA and Linac$%
\otimes $LHC based $\gamma p$ colliders are shown in Figures \ref{fig6} and 
\ref{fig7}.

In Eq. (\ref{Eq.25}), $f_{\gamma /e}(y)$ is the function which describes the
spectrum of photons scattered backward from the interaction of laser light
with the high energy electron beam (\ref{Eq.23}) or the Weizsaecker-Williams
(WW) approximated photon spectrum \cite{9}

\begin{equation}
f_{\gamma /e}(y)=\frac{\alpha }{2\pi }\log \left[ \frac{(1-y)}{y^{2}\delta }%
\left( \frac{1+(1-y)^{2}}{y}\right) -\frac{2(1-y-\delta y^{2})}{y}\right]
\end{equation}
where $\alpha $ is the fine structure constant, $\delta =(m/Q_{\max })^{2},$ 
$Q_{\max }$ being the photon virtuality region, and $m$ is the mass of
incoming particle. It should be noted that the total cross sections with
Compton bacscattered photons are about ten times larger than the
corresponding cross sections with the WW photons. This makes the $\gamma p$
colliders powerful machines in searching for the new physics. The total
number of signal events are given in Table \ref{table3}.

In conclusion, top quark can be produced at $\gamma p$ colliders in the
resonance channel via anomalous interaction. The main decay mode for $t$
quarks is the $b+W.$ For this channel $b$ quarks can be identified in the
detector as $b-$jets (so called b-tagging), the hadronic decay modes of $W-$%
boson will be identified as two-jets and its leptonic decay as a
lepton+missing $p_{T}.$ The $b-$quarks from the decay of top quarks have
higher transverse momentum than those from the backgrounds. This makes the
signal separable from the backgrounds even at small $f$. As can be seen from
the Tables \ref{table2} and \ref{table3}, we can observe the anomalous
interactions for $t$ quarks down to the couplings $f=2\times 10^{-3}$ for
THERA and $f=2\times 10^{-4}$ for Linac$\otimes $LHC based $\gamma p$
colliders within the statistical acceptance.

\newpage

\begin{table}[tbp]
\caption{Main parameters of the THERA and Linac$\otimes $LHC based $\protect%
\gamma p$ colliders }
\label{table1}
\begin{tabular}{ccccc}
\hline
Machine & $E_{e}$(GeV) & $E_{p}$(GeV) & $\sqrt{s_{\gamma p}^{\max }}$(GeV) & 
$L_{\gamma p}($cm$^{-2}$s$^{-1})$ \\ \hline
THERA$
\begin{array}{c}
1 \\ 
2 \\ 
3
\end{array}
$ & $
\begin{array}{c}
250 \\ 
500 \\ 
800
\end{array}
$ & $
\begin{array}{c}
1000 \\ 
500 \\ 
800
\end{array}
$ & $
\begin{array}{c}
911 \\ 
911 \\ 
1456
\end{array}
$ & $
\begin{array}{c}
4\times 10^{30} \\ 
2.5\times 10^{31} \\ 
1.6\times 10^{31}
\end{array}
$ \\ \hline
Linac$\otimes $LHC & $1000$ & $7000$ & $4820$ & $1\times 10^{33}$ \\ \hline
\end{tabular}
\end{table}

\begin{table}[tbp]
\caption{Cross sections in the chosen $p_{T}$ window (50-70 GeV) and
statistical significance for single top quark production at the THERA and
Linac$\otimes $LHC based $\protect\gamma p$ colliders. S and B stand for the
number of events for signal and background, respectively. }
\begin{tabular}{llllll}
\hline
Machine $\downarrow $ & f$\rightarrow $ & $10^{-1}$ & $10^{-2}$ & $10^{-3}$
& $10^{-4}$ \\ \hline
THERA 1 & $
\begin{array}{c}
\Delta \sigma ^{\text{S}}(\text{pb}) \\ 
\text{S}/\sqrt{\text{S+B}}
\end{array}
$ & $
\begin{array}{c}
7.94\times 10^{1} \\ 
9.76\times 10^{1}
\end{array}
$ & $
\begin{array}{c}
1.24\times 10^{0} \\ 
1.22\times 10^{1}
\end{array}
$ & $
\begin{array}{c}
1.24\times 10^{-2} \\ 
1.18\times 10^{0}
\end{array}
$ & $
\begin{array}{c}
1.24\times 10^{-4} \\ 
4.\,\allowbreak 28\times 10^{-2}
\end{array}
$ \\ \hline
THERA 2 & $
\begin{array}{c}
\Delta \sigma ^{\text{S}}(\text{pb}) \\ 
\text{S}/\sqrt{\text{S+B}}
\end{array}
$ & $
\begin{array}{c}
7.94\times 10^{1} \\ 
2.44\times 10^{2}
\end{array}
$ & $
\begin{array}{c}
1.24\times 10^{0} \\ 
3.05\times 10^{1}
\end{array}
$ & $
\begin{array}{c}
1.24\times 10^{-2} \\ 
2.95\times 10^{0}
\end{array}
$ & $
\begin{array}{c}
1.24\times 10^{-4} \\ 
1.07\times 10^{-1}
\end{array}
$ \\ \hline
THERA 3 & $
\begin{array}{c}
\Delta \sigma ^{\text{S}}(\text{pb}) \\ 
\text{S}/\sqrt{\text{S+B}}
\end{array}
$ & $
\begin{array}{c}
4.27\times 10^{1} \\ 
1.43\times 10^{2}
\end{array}
$ & $
\begin{array}{c}
6.66\times 10^{-1} \\ 
1.78\times 10^{1}
\end{array}
$ & $
\begin{array}{c}
6.69\times 10^{-3} \\ 
1.52\times 10^{0}
\end{array}
$ & $
\begin{array}{c}
6.69\times 10^{-5} \\ 
\allowbreak 2.82\times 10^{-2}
\end{array}
$ \\ \hline
Linac$\otimes $LHC & $
\begin{array}{c}
\Delta \sigma ^{\text{S}}(\text{pb}) \\ 
\text{S}/\sqrt{\text{S+B}}
\end{array}
$ & $
\begin{array}{c}
1.56\times 10^{2} \\ 
2.16\times 10^{3}
\end{array}
$ & $
\begin{array}{c}
2.43\times 10^{0} \\ 
2.69\times 10^{2}
\end{array}
$ & $
\begin{array}{c}
2.44\times 10^{-2} \\ 
2.06\times 10^{1}
\end{array}
$ & $
\begin{array}{c}
2.45\times 10^{-4} \\ 
\,\allowbreak 3.15\times 10^{-1}
\end{array}
$ \\ \hline
\end{tabular}
\label{table2}
\end{table}

\begin{table}[tbp]
\caption{Total cross sections and number of events for single top quark
production. N$_{i}$ are the number of signal events for the integrated
luminosities given in the text.}
\label{table3}
\begin{tabular}{lllll}
\hline
$f\rightarrow $ & $10^{-1}$ & $10^{-2}$ & $10^{-3}$ & $10^{-4}$ \\ \hline
$\sigma _{1,2}($pb$)$ & $1.13\times 10^{2}$ & $1.27\times 10^{0}$ & $%
1.27\times 10^{-2}$ & $1.27\times 10^{-4}$ \\ \hline
N$_{1}$ & $1.36\times 10^{4}$ & $1.52\allowbreak \times 10^{2}$ & $%
1.52\times 10^{0}$ & $1.52\times 10^{-2}$ \\ \hline
N$_{2}$ & $8.48\times 10^{4}$ & $9.53\times 10^{2}$ & $9.53\times 10^{0}$ & $%
9.53\times 10^{-2}$ \\ \hline
$\sigma _{3}($pb$)$ & $7.87\times 10^{1}$ & $7.99\times 10^{-1}$ & $%
7.99\times 10^{-3}$ & $7.99\times 10^{-5}$ \\ \hline
N$_{3}$ & $3.78\times 10^{4}$ & $3.83\,\allowbreak \times 10^{2}$ & $%
3.\,\allowbreak 84\times 10^{0}$ & $\,\allowbreak 3.83\times 10^{-2}$ \\ 
\hline
$\sigma _{\text{LHC}}($pb$)$ & $2.31\times 10^{2}$ & $2.31\times 10^{0}$ & $%
2.31\times 10^{-2}$ & $2.31\times 10^{-4}$ \\ \hline
N$_{\text{LHC}}$ & $6.93\times 10^{6}$ & $6.93\times 10^{4}$ & $6.93\times
10^{2}$ & $6.93\times 10^{0}$ \\ \hline
\end{tabular}
\end{table}

\newpage

\begin{center}
\begin{figure}[tbp]
\epsfig{file=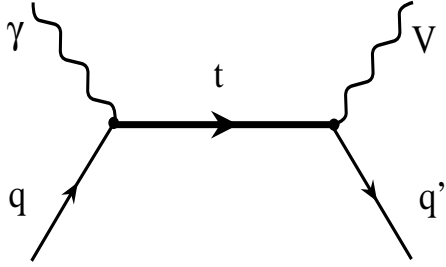,width=6.5cm,height=4.5cm}
\caption{Feynman diagram for single $t$ production. Here q denotes the
quarks $u$ or $c$; V stands for the gauge bosons $(\protect\gamma ,Z,g,W).$ }
\label{fig1}
\end{figure}

\vspace{1cm}

\begin{figure}[tbp]
\epsfig{file=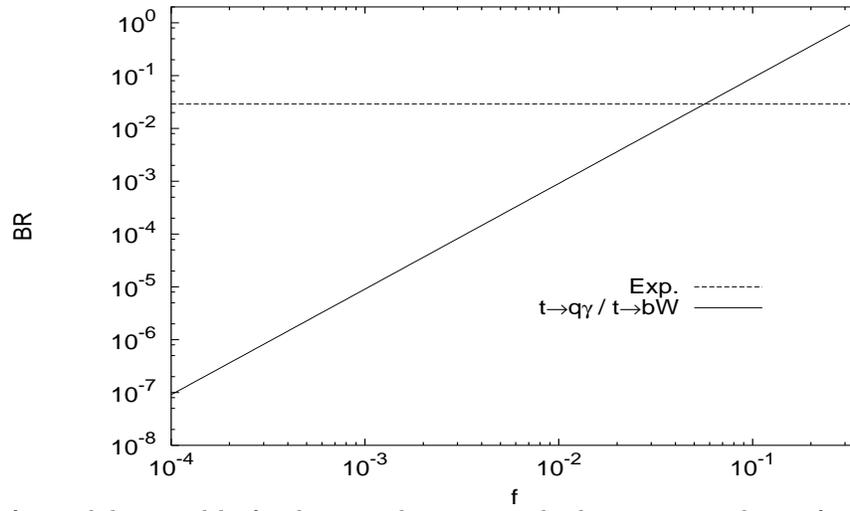,width=12cm,height=7cm}
\caption{The ratio of partial decay widths for the anomalous top quark, the
experimental ratio for this channel is also given, $f=|f_{1}|=|f_{2}|$. }
\label{fig2}
\end{figure}

\vspace{1cm}

\begin{figure}[tbp]
\epsfig{file=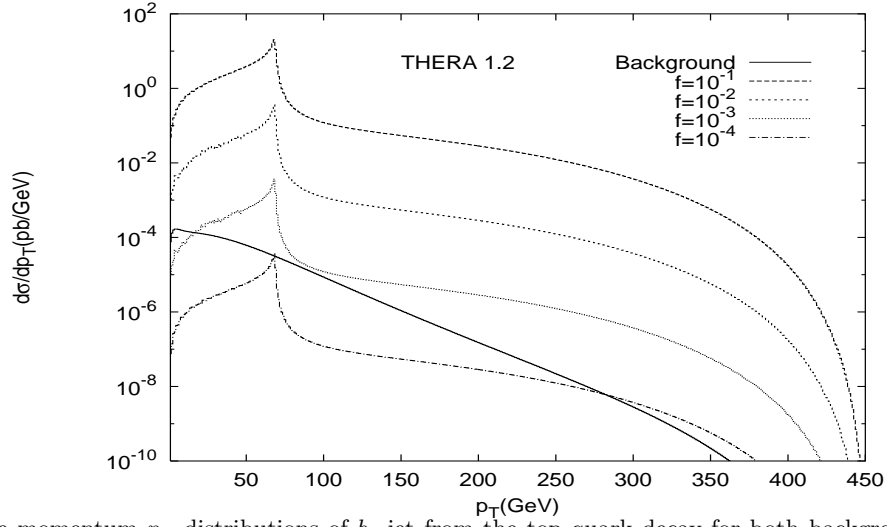,width=12cm,height=7cm}
\caption{Transverse momentum $p_{T}$ distributions of $b-$jet from the top
quark decay for both background and the signal at $f=0.1,$ $0.01,$ $0.001$
and $0.0001.$ }
\label{fig3}
\end{figure}

\vspace{1cm}

\begin{figure}[tbp]
\epsfig{file=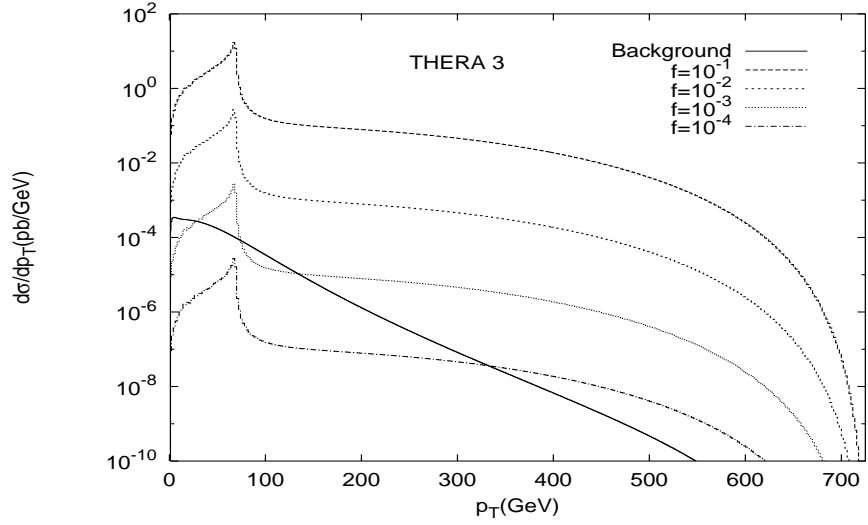,width=12cm,height=7cm}
\caption{The same as Figure 3 for THERA 3 option.}
\label{fig4}
\end{figure}

\vspace{1cm}

\begin{figure}[tbp]
\epsfig{file=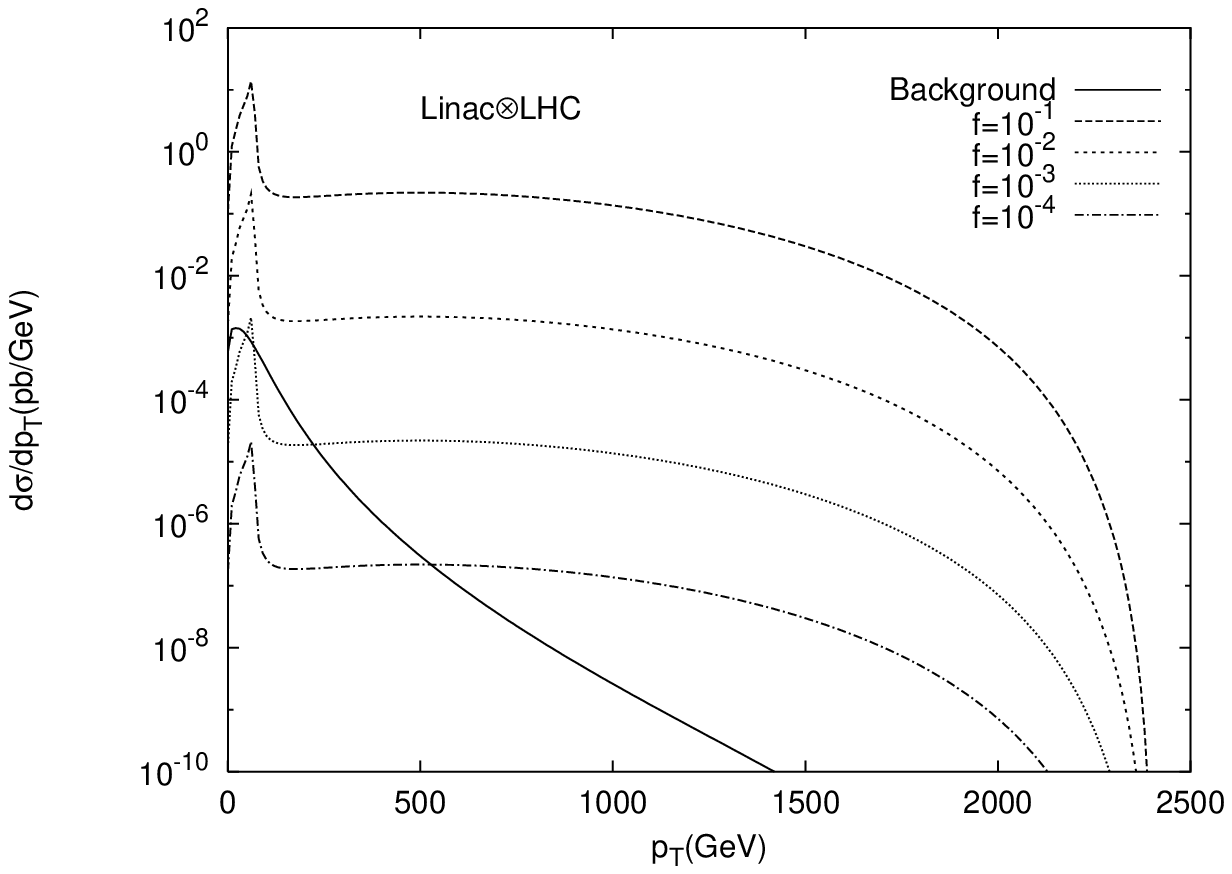,width=12cm,height=7cm}
\caption{The same as Figure 3 for Linac$\otimes $LHC based $\protect\gamma p$
collider.}
\label{fig5}
\end{figure}

\vspace{1cm}

\begin{figure}[tbp]
\epsfig{file=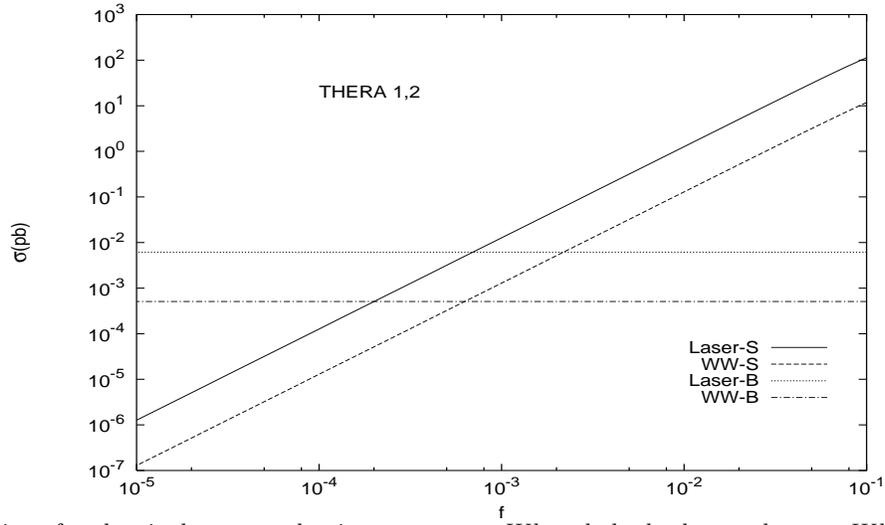,width=12cm,height=7cm}
\caption{Cross sections for the single top production $\protect\gamma %
p\rightarrow t\rightarrow Wb$ and the background $\protect\gamma %
p\rightarrow Wb$ using laser and WW photon spectrum depending on the
coupling $f$ at THERA 1 and 2 based $\protect\gamma p$ colliders.}
\label{fig6}
\end{figure}

\vspace{1cm}

\begin{figure}[tbp]
\epsfig{file=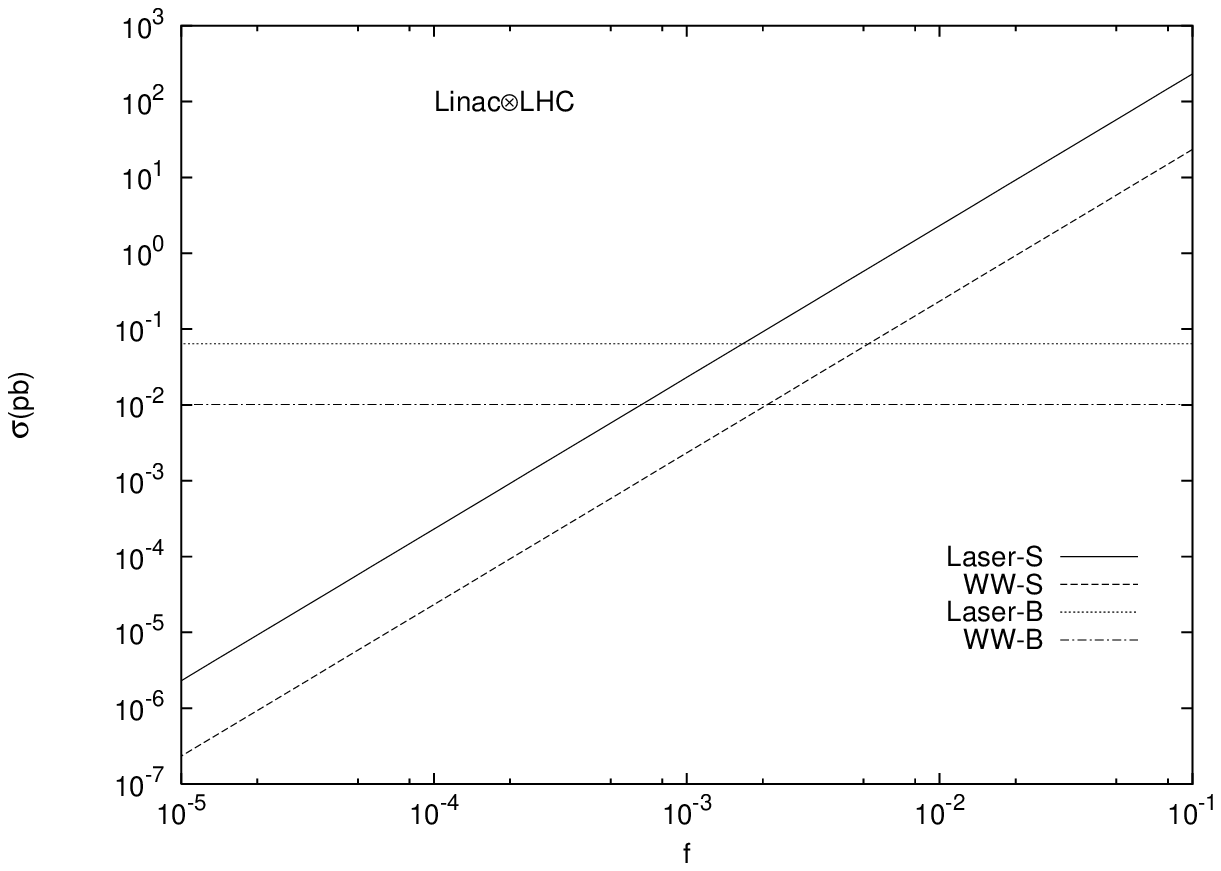,width=12cm,height=7cm}
\caption{The same as Figure 6 for Linac$\otimes $LHC based $\protect\gamma p$
collider.}
\label{fig7}
\end{figure}
\end{center}

\end{document}